{}{}%
\newenvironment{centre}{\begin{center}}{\end{center}}\documentstyle[12pt]{article}%
\newlength\Cscr\newlength\Csave\newlength\Ctenthex\newlength\CFxsize\newlength\CFxsizeps\newlength
\CFsizemakebox\newlength\CFleftcrop\newlength\CFrightcrop
\newlength\CZtbldist\newlength\CZfigdist\newlength\CGDnum\newlength\CGDtext
\newcounter{Cscr}\newcounter{CBcit}\newcounter{CTle%
tter}\newcounter{Ceqindent}\newcounter{C%
Btnc}\newcounter{CBtntc}\newcounter{CEht}\newcounter{CbsA}%
\newcounter{CbsB}\newcounter{CbsC}\newcounter{CbsD}\hoffset\Cscr
\voffset\Cscr\protect
\begin{document}%
\renewcommand\theequation{\arabic{equation}}\renewcommand
\thetable{\arabic{table}}\renewcommand\thefigure{\arabic{figure}%
}\renewcommand\thesection{\Roman{section}}\renewcommand
\thesubsection{\Alph{subsection}}\renewcommand\thesubsubsection{%
\arabic{subsubsection}}\setcounter{CEht}{10}\setcounter{CbsA}{1}%
\setcounter{CbsB}{1}\setcounter{CbsC}{1}\setcounter{CbsD}{1}%
\setlength{\CFxsize}{0.998\hsize}\hfill UM--P--95/12\par\hfill R%
CHEP--95/08\par\hfill To appear in the \mbox{}\protect\/{\protect
\em American Journal of Physics\protect\/}\par\addtocounter{CBtn%
tc}{1}{\centering\protect\mbox{}\\*[\baselineskip]{\large\bf The
Foldy--Wouthuysen transformation$^{\fnsymbol{CBtnc}}$%
\addtocounter{CBtnc}{1}}\\*}\addtocounter{CBtntc}{1}\addtocounter
{CBtntc}{1}{\centering\protect\mbox{}\\John P.~Costella$^{%
\fnsymbol{CBtnc}}$\addtocounter{CBtnc}{1} and Bruce H.~J.~McKell%
ar$^{\fnsymbol{CBtnc}}$\addtocounter{CBtnc}{1}\\*}{\centering{%
\small\mbox{}\protect\/{\protect\em School of Physics, The Unive%
rsity of Melbourne, Parkville, Victoria 3052, Australia\protect
\/}}\\}{\centering\protect\mbox{}\\(23 March 1995)\\} \par\vspace
\baselineskip\begin{centre}{\small\bf Abstract}\end{centre}%
\vspace{-1.5ex}\vspace{-0.75\baselineskip}\par\setlength{\Csave}%
{\parskip}\begin{quote}\setlength{\parskip}{0.5\baselineskip}%
\small\noindent The Foldy--Wouthuysen transformation of the Dira%
c Hamiltonian is generally taught as simply a mathematical trick
that allows one to obtain a two-component theory in the low-ener%
gy limit. It is not often emphasized that the transformed repres%
entation is the only one in which one can take a meaningful \mbox
{}\protect\/{\protect\em classical limit\protect\/}, in terms of
particles and antiparticles. We briefly review the history and p%
hysics of this transformation. \end{quote}\setlength{\parskip}{%
\Csave}\par\refstepcounter{section}\vspace{1.5\baselineskip}\par
{\centering\bf\thesection. Introduction\\*[0.5\baselineskip]}%
\protect\indent\label{sect:Intro}In relativistic quantum mechani%
cs, one often identifies those components of the wavefunction th%
at represent ``particles'', and those that represent ``antiparti%
cles''. But there always exist \mbox{}\protect\/{\protect\em can%
onical transformations\protect\/} of the wavefunction (changes o%
f representation) that mix these particle and antiparticle compo%
nents together, while still leaving the physical quantities repr%
esented by the theory unchanged, as long as the operators are co%
mplementarily transformed. This means that the components of the
wavefunction that appear to represent antiparticles in one repre%
sentation will actually be a \mbox{}\protect\/{\protect\em super%
position\protect\/} of particle and antiparticle components in a
different representation.\par It would be difficult to recognize
a classical limit of the relativistic quantum theory if this arb%
itrariness in representation were to be permitted to run free. C%
lassical physics does not have any trouble with the concept of a%
ntiparticles per~se: by Feynman's interpretation, antiparticle m%
otion is simply effected by means of the ``classical $C$'' trans%
formation \setcounter{Ceqindent}{0}\protect\begin{eqnarray}\tau
\rightarrow-\tau\protect\nonumber\setlength{\Cscr}{\value{CEht}%
\Ctenthex}\addtolength{\Cscr}{-1.0ex}\protect\raisebox{0ex}[%
\value{CEht}\Ctenthex][\Cscr]{}\protect\end{eqnarray}\setcounter
{CEht}{10}on the corresponding particle motion, where $\tau$ is
the proper-time of the particle. But the discreteness of this cl%
assical $C$ transformation---and the lack of any sort of ``super%
position'' principle---mean that classical physics does \mbox{}%
\protect\/{\protect\em not\protect\/} admit any ``mixing'' of pa%
rticle and antiparticle motion.\par\refstepcounter{section}%
\vspace{1.5\baselineskip}\par{\centering\bf\thesection. Newton a%
nd Wigner\\*[0.5\baselineskip]}\protect\indent\label{sect:NW}How%
, then, can one obtain a meaningful classical limit of relativis%
tic quantum mechanics? The clue to the path out of this dilemma
was first found in 1949 by \protect\ref{au:Newton1949},$^{\ref{c%
it:Newton1949}}$ as almost a by-product of other, more abstract
considerations. The findings of Newton and Wigner eradicated som%
e of the myths surrounding the \mbox{}\protect\/{\protect\em pos%
ition operator\protect\/} in relativistic wave equations---in pa%
rticular, that states localized in position cannot be formed sol%
ely from positive-energy states; and that if a particle's positi%
on is measured below its Compton wavelength, one necessarily gen%
erates particle--antiparticle pairs, which renders the position
measurement of a single particle impossible. In pursuing some ra%
ther simple questions of a group theoretical nature, they not on%
ly found what they were looking for, but also some unexpected bo%
nuses. These were explained and elaborated on by \protect\ref{au%
:Foldy1950},$^{\ref{cit:Foldy1950}}$ who also obtained the expli%
cit transformation that realized the goals of \protect\ref{au:Ne%
wton1949} for the physically important case of a spin-half parti%
cle. (\protect\ref{au:Case1954}$^{\ref{cit:Case1954}}$ later gen%
eralized their method to spin-zero and spin-one particles.)\par T%
he original aim of Newton and Wigner was to rigorously formulate
the properties of \mbox{}\protect\/{\protect\em localized states%
\protect\/}, for arbitrary-spin relativistic representations of
elementary particles. They proceeded simply on the basis of \mbox
{}\protect\/{\protect\em invariance requirements\protect\/}. The%
y sought a set of states which were localized at a certain point
in space, such that any state becomes, after a translation, orth%
ogonal to all of the undisplaced states; such that the superposi%
tion of any two such localized states is again a localized state
in the set; that the set of states be invariant under rotations
about the point of localization, and under temporal and spatial
reflections; and that the states all satisfy certain regularity
conditions, amounting to the requirement that all of the operato%
rs of the Lorentz group be applicable to them.\par{}From such a
simple and reasonable set of requirements, a most bountiful crop
was harvested. Firstly, Newton and Wigner found that the set of
states they sought \mbox{}\protect\/{\protect\em could\protect\/%
}, indeed, be found, for arbitrary spin (provided the mass is no%
n-zero); moreover, their requirements in fact specify a \mbox{}%
\protect\/{\protect\em unique\protect\/} set of states with the
desired properties. Furthermore, these states are all \mbox{}%
\protect\/{\protect\em purely positive-energy states\protect\/}
(or, equivalently, purely negative-energy). They further belong
to a \mbox{}\protect\/{\protect\em continuous eigenvalue spectru%
m of a particular operator\protect\/}, which itself has the prop%
erty of preserving the positive-energy nature of the wavefunctio%
n.\par Due to these remarkably agreeable properties, Newton and
Wigner felt that one would be justified in referring to the oper%
ator they had found as \mbox{}\protect\/{\protect\em the\protect
\/} position operator---in contradistinction to the operator ${%
\protect\mbox{\protect\boldmath{$x$}}}$ in some arbitrary repres%
entation of the relativistic wave equation, which only is the ``%
position'' operator \mbox{}\protect\/{\protect\em in that partic%
ular representation\protect\/}, and hence has no invariant physi%
cal meaning, since the representation may be subject to an (in g%
eneral position-dependent) canonical transformation, that by def%
inition cannot change any physical quantities, but which most de%
finitely changes the expectation values of the fixed operator ${%
\protect\mbox{\protect\boldmath{$x$}}}$. The Newton--Wigner posi%
tion operator had, in fact, been discovered previously in 1935 b%
y \protect\ref{au:Pryce1935},$^{\ref{cit:Pryce1935}}$ who found
the operator a useful tool in the Born--Infeld theory, and, late%
r,$^{\ref{cit:Pryce1948}}$ in a discussion of relativistic defin%
itions of the center of mass for systems of particles.\par
\refstepcounter{section}\vspace{1.5\baselineskip}\par{\centering
\bf\thesection. Foldy and Wouthuysen\\*[0.5\baselineskip]}%
\protect\indent\label{sect:FW}A natural question to ask, given t%
he findings of Newton and Wigner, is the following: What does a
given relativistic wave equation look like in the representation
in which the Newton--Wigner position operator \mbox{}\protect\/{%
\protect\em is\protect\/}, in fact, simply the vector ${\protect
\mbox{\protect\boldmath{$x$}}}$? This is the question effectivel%
y asked by Foldy and Wouthuysen in their classic 1950 paper,$^{%
\ref{cit:Foldy1950}}$ for the physically important case of the D%
irac equation. (Their stated aim was actually to find a represen%
tation in which the components for positive- and negative-energy
states are decoupled, but from the above it is clear that this i%
s effectively the same as seeking the Newton--Wigner representat%
ion.) What they found is, even today, simply astounding. Firstly%
, they found that the canonical transformation from the Dirac--P%
auli representation to the Newton--Wigner representation of the
\mbox{}\protect\/{\protect\em free\protect\/} Dirac equation is,
in fact, obtainable exactly. Secondly, they found that the \mbox
{}\protect\/{\protect\em Hamiltonian\protect\/} for the free par%
ticle, in the Newton--Wigner representation, agrees completely w%
ith that of classical physics, \setcounter{Ceqindent}{0}\protect
\begin{eqnarray}\protect\left.\protect\begin{array}{rcl}\protect
\displaystyle H_{\mbox{\scriptsize NW}}=\beta(m^2+{\protect\mbox
{\protect\boldmath{$p$}}}^2)^{1/2}\equiv\beta W_p\setlength{\Cscr
}{\value{CEht}\Ctenthex}\addtolength{\Cscr}{-1.0ex}\protect
\raisebox{0ex}[\value{CEht}\Ctenthex][\Cscr]{}\protect\end{array%
}\protect\right.\protect\label{eq:FW-HamNW}\protect\end{eqnarray%
}\setcounter{CEht}{10}(we use units in which $\hbar=c=1$), in co%
ntrast to that of the Dirac--Pauli representation, \setcounter{C%
eqindent}{0}\protect\begin{eqnarray}H_{\mbox{\scriptsize DP}}=%
\beta m+{\protect\mbox{\protect\boldmath{$\alpha$}}}\mbox{$%
\hspace{0.2ex}\cdot\hspace{0.2ex}$}{\protect\mbox{\protect
\boldmath{$p$}}},\protect\nonumber\setlength{\Cscr}{\value{CEht}%
\Ctenthex}\addtolength{\Cscr}{-1.0ex}\protect\raisebox{0ex}[%
\value{CEht}\Ctenthex][\Cscr]{}\protect\end{eqnarray}\setcounter
{CEht}{10}which---while having the important property of lineari%
ty---does not resemble the classical expression at all.\par Thir%
dly, Foldy and Wouthuysen found that the \mbox{}\protect\/{%
\protect\em velocity operator\protect\/} (obtained from the posi%
tion operator by means of its Heisenberg equation of motion) in
the Newton--Wigner representation---or, equivalently, the corres%
ponding Newton--Wigner velocity operator in \mbox{}\protect\/{%
\protect\em any\protect\/} representation---satisfies the \mbox{%
}\protect\/{\protect\em classical\protect\/} relation for a free
particle: \setcounter{Ceqindent}{0}\protect\begin{eqnarray}%
\protect\left.\protect\begin{array}{rcl}\protect\displaystyle{%
\protect\mbox{\protect\boldmath{$v$}}}_{\mbox{\scriptsize NW}}%
\equiv\mbox{$\protect\displaystyle\protect\frac{d}{dt}$}{\protect
\mbox{\protect\boldmath{$x$}}}_{\mbox{\scriptsize NW}}=\beta\mbox
{$\protect\displaystyle\protect\frac{{\protect\mbox{\protect
\boldmath{$p$}}}}{W_p}$}. \setlength{\Cscr}{\value{CEht}\Ctenthex
}\addtolength{\Cscr}{-1.0ex}\protect\raisebox{0ex}[\value{CEht}%
\Ctenthex][\Cscr]{}\protect\end{array} \protect\right.\protect
\label{eq:FW-VNW}\protect\end{eqnarray}\setcounter{CEht}{10}That
(\protect\ref{eq:FW-VNW}) is an amazing result is recognized fro%
m the fact that, from the very inception of the Dirac equation,
it was known that the ``velocity'' operator in the \mbox{}%
\protect\/{\protect\em Dirac--Pauli\protect\/} representation
\mbox{}\protect\/{\protect\em does not\protect\/} make any class%
ical sense whatsoever: its sole eigenvalues are plus or minus th%
e speed of light; it is not directly related to the momentum ${%
\protect\mbox{\protect\boldmath{$p$}}}$; and its equation of mot%
ion has non-real ``\mbox{}\protect\/{\protect\em zitterbewegung%
\protect\/}'' oscillatory motion (see, e.g., Ref.~\ref{cit:Dirac%
1958}). In drastic contradistinction, the Newton--Wigner velocit%
y operator ${\protect\mbox{\protect\boldmath{$v$}}}_{\mbox{%
\scriptsize NW}}$ of (\protect\ref{eq:FW-VNW}) has the physicall%
y understandable continuum of eigenvalues between plus and minus
the speed of light; its relationship to the momentum of the free
particle is identical to that valid in classical physics; and, w%
hen one considers in turn \mbox{}\protect\/{\protect\em its%
\protect\/} Heisenberg equation of motion, then one finds that,
for a free particle, the velocity ${\protect\mbox{\protect
\boldmath{$v$}}}_{\mbox{\scriptsize NW}}$ is a constant, since $%
{\protect\mbox{\protect\boldmath{$p$}}}$ and $W$ are also.\par F%
ourthly, Foldy and Wouthuysen found that the free-particle spin
and orbital angular momentum operators in the Newton--Wigner rep%
resentation---defined to be simply \mbox{${\protect\mbox{\protect
\boldmath{$l$}}}_{\mbox{\scriptsize NW}}\equiv{\protect\mbox{%
\protect\boldmath{$x$}}}\mbox{$\times$}{\protect\mbox{\protect
\boldmath{$p$}}}$} and \mbox{${\protect\mbox{\protect\boldmath{$%
\sigma$}}}_{\mbox{\scriptsize NW}}\equiv{\protect\mbox{\protect
\boldmath{$\sigma$}}}$} in this representation---are \mbox{}%
\protect\/{\protect\em constants of the motion separately\protect
\/}; again, it is well-known$^{\ref{cit:Dirac1958}}$ that, in th%
e Dirac--Pauli representation, these operators are \mbox{}%
\protect\/{\protect\em not\protect\/} separately constants of th%
e motion, even for a free particle. (The peculiarity of the Dira%
c--Pauli representation in this respect can be traced back to th%
e fact that the ``position'' operator in that representation exh%
ibits the non-physical ``\mbox{}\protect\/{\protect\em zitterbew%
egung\protect\/}'' motion, which thus enters into the motion of
the ``orbital angular momentum'' operator \mbox{${\protect\mbox{%
\protect\boldmath{$x$}}}_{\mbox{\scriptsize DP}}\mbox{$\times$}{%
\protect\mbox{\protect\boldmath{$p$}}}$} in this representation.%
)\par As a fifth and final accomplishment, Foldy and Wouthuysen
attacked the problem of finding the canonical transformation fro%
m the Dirac--Pauli representation to the Newton--Wigner represen%
tation, in the case of the \mbox{}\protect\/{\protect\em elec\-t%
ro\-mag\-net\-ic\-ally-coupled\protect\/} Dirac equation. Unfort%
unately, this cannot be done in closed form. Nevertheless, Foldy
and Wouthuysen showed how one can obtain successive approximatio%
ns to the required transformation, as a power series in \mbox{$p%
^\alpha\!/m$}, \mbox{$qA^\alpha\!/m$}, for an arbitrary initial
Hamiltonian $H_{\mbox{\scriptsize DP}}$ in the Dirac--Pauli repr%
esentation; this is the transformation that is presented in almo%
st any textbook on relativistic quantum mechanics (see, e.g., Re%
f.~\ref{cit:Bjorken1964}).\par An unstated assumption, crucial t%
o the validity of the power series implementation of the transfo%
rmation, is that the ``odd'' part of the Hamiltonian is in fact
of no higher order in $m$ than $m^0$. This is usually the case,
but the assumption has the latent ability to trip one up. For ex%
ample, if one tries to transform a Hamiltonian in which the mass
term $\beta m$ has been multiplied by \mbox{$e^{i\theta\gamma_5}%
$}---say, by a canonical transformation of the representation,%
---then one can be led to quite erroneous conclusions if one ass%
umes that the terms omitted in the subsequent power series Foldy%
--Wouthuysen expansion are of high order in \mbox{$p^\alpha\!/m$%
} and \mbox{$qA^\alpha\!/m$}; in fact, the omitted terms are of
exactly the same order as the terms that are retained; the power
series method is, if applied in this way, completely useless. In
such cases, the correct procedure is to first perform a simple c%
anonical transformation to remove the order \mbox{$m^{+1}$} term%
s from the ``odd'' parts of the Hamiltonian; the resulting repre%
sentation may then be fruitfully subjected to the power series F%
oldy--Wouthuysen transformation.\par\refstepcounter{section}%
\vspace{1.5\baselineskip}\par{\centering\bf\thesection. The Dira%
c--Pauli representation\\*[0.5\baselineskip]}\protect\indent
\label{sect:DP}It may be wondered, after hearing of all of the w%
onderful properties of the Newton--Wigner representation, why on%
e should bother with any other representation at all. In particu%
lar, why do we usually only concentrate on the Dirac--Pauli repr%
esentation of the Dirac equation? (Or representations ``triviall%
y'' related to it; we shall define this term with more precision
shortly.) The answer is subtle, but beautiful. \mbox{}\protect\/%
{\protect\em The charged leptons in Nature are well described by
a minimal coupling of their Dirac fields to the electromagnetic
field, in the Dirac--Pauli representation only.\protect\/} It is
not often stressed that \mbox{}\protect\/{\protect\em minimal co%
upling\protect\/}---the use of the prescription \setcounter{Ceqi%
ndent}{0}\protect\begin{eqnarray}p\rightarrow p-qA\protect
\nonumber\setlength{\Cscr}{\value{CEht}\Ctenthex}\addtolength{%
\Cscr}{-1.0ex}\protect\raisebox{0ex}[\value{CEht}\Ctenthex][\Cscr
]{}\protect\end{eqnarray}\setcounter{CEht}{10}in the correspondi%
ng non-interacting formalism---is \mbox{}\protect\/{\protect\em n%
ot\protect\/} a universal, rep\-re\-sen\-tation-indep\-endent tr%
ansformation. The reason is that, in general, a canonical transf%
ormation used to effect a change in representation may be \mbox{%
}\protect\/{\protect\em momentum-dependent\protect\/}; indeed, t%
he Foldy--Wouthuysen transformation itself is an important examp%
le. Clearly, the processes of using minimal coupling, and then p%
erforming a momentum-dependent transformation, on the one hand;
and that of performing the momentum-dependent transformation fir%
st, and \mbox{}\protect\/{\protect\em then\protect\/} using mini%
mal coupling, on the other; will lead to completely \mbox{}%
\protect\/{\protect\em different\protect\/} relativistic wave eq%
uations, in general. \mbox{}\protect\/{\protect\it A~priori,
\protect\/}\mbox{}\protect\/{\protect\em one cannot know which r%
epresentation one should use the minimal coupling prescription o%
n.\protect\/}\par(Clearly, ``trivial'' changes of representation%
, in the sense used above, are therefore those in which the cano%
nical transformation does not involve the momentum operator.)\par
We thus see that, by his insistence on a \mbox{}\protect\/{%
\protect\em linear\protect\/} relationship between ${\protect
\mbox{\protect\boldmath{$p$}}}$ and $H$---for reasons that were
rendered obsolete by second quantization---Dirac was led to the
one representation of the spin-half Hamiltonian in which the ass%
umption of minimal coupling gives the correct electromagnetic in%
teractions for the electron, and in particular the correct gyrom%
agnetic ratio and hydrogen spectrum. With hindsight, we can see
that Dirac was both brilliant and lucky.\par\refstepcounter{sect%
ion}\vspace{1.5\baselineskip}\par{\centering\bf\thesection. The
two faces of the electron\\*[0.5\baselineskip]}\protect\indent
\label{sect:TwoFaces}We therefore come to recognize that there a%
re \mbox{}\protect\/{\protect\em two\protect\/} representations
of the Dirac equation that are singled out above all others---ea%
ch having qualities unique to itself---that have a truly direct
correspondence with Nature: The Dirac--Pauli representation is u%
nique due to its linearity; it is the representation in which th%
e charged leptons are minimally coupled. The Newton--Wigner repr%
esentation is unique due to its decoupling of positive- and nega%
tive-energy states; it is the representation in which the operat%
ors of the theory correspond to their classical counterparts.\par
We may go even further, conceptually speaking, in our descriptio%
n of the charged leptons: they are, in effect, two types of part%
icle in the one being. On the one hand, in the Dirac--Pauli repr%
esentation, all four components are inextricably coupled, but th%
e particles are \mbox{}\protect\/{\protect\em pure, pointlike, s%
tructureless electric charges\protect\/}. On the other hand, in
the Newton--Wigner representation, operators act quite in accord
with classical mechanics, but the electromagnetic interactions a%
re more complicated: they still have electric charge, but throug%
h the Foldy--Wouthuysen transformation they acquire a \mbox{${%
\protect\mbox{\protect\boldmath{$\mu$}}}\mbox{$\hspace{0.2ex}%
\cdot\hspace{0.2ex}$}{\protect\mbox{\protect\boldmath{$B$}}}$}
\mbox{}\protect\/{\protect\em magnetic moment\protect\/} interac%
tion, and (less well-known) an \mbox{}\protect\/{\protect\em ele%
ctric charge radius\protect\/} (manifested in the ``Darwin term%
'' in the Hamiltonian; see Refs.~\ref{cit:Foldy1951}, \ref{cit:F%
oldy1952a}, \ref{cit:Foldy1952b}, \ref{cit:Foldy1958}).\par
\vspace{1.5\baselineskip}\par{\centering\bf Acknowledgments\\*[0%
.5\baselineskip]}\protect\indent Helpful discussions with I.~Khr%
iplovich, J.~Anandan, G.~I.~Opat, J.~W.~G.~Wig\-nall, A.~J.~Da\-%
vies, M.~J.~Thom\-son, T.~D.~Kieu, D.-D.~Wu, and S.~Bass are gra%
tefully acknowledged. This work was supported in part by the Aus%
tralian Research Council, an Australian Postgraduate Research Al%
lowance and a Dixson Research Scholarship. We warmly thank the I%
nstitute for Nuclear Theory at the University of Washington for
its hospitality and the United States Department of Energy Grant~%
\#DOE/ER40561 for partial support during the progress of this w%
ork.\par\vspace{1.5\baselineskip}\par{\centering\bf References%
\\*[0.5\baselineskip]}\protect\mbox{}\vspace{-\baselineskip}%
\vspace{-2ex}\settowidth\CGDnum{[\ref{citlast}]}\setlength{%
\CGDtext}{\textwidth}\addtolength{\CGDtext}{-\CGDnum}\begin{list%
}{Error!}{\setlength{\labelwidth}{\CGDnum}\setlength{\labelsep}{%
0.75ex}\setlength{\leftmargin}{0ex}\setlength{\rightmargin}{0ex}%
\setlength{\itemsep}{0ex}\setlength{\parsep}{0ex}}\protect
\frenchspacing\setcounter{CBtnc}{1}\item[{\hfill\makebox[0ex][r]%
{\raisebox{0ex}[1ex][0ex]{$^{\mbox{$\fnsymbol{CBtnc}$}}$}}}]%
\addtocounter{CBtnc}{1}This paper has been taken from Section~4.%
4.1 of JPC's Ph.D.\ thesis.$^{\ref{cit:Costella1994}}$\item[{%
\hfill\makebox[0ex][r]{\raisebox{0ex}[1ex][0ex]{$^{\mbox{$%
\fnsymbol{CBtnc}$}}$}}}]\addtocounter{CBtnc}{1}E-mail address: j%
pc@physics.unimelb.edu.au.\item[{\hfill\makebox[0ex][r]{\raisebox
{0ex}[1ex][0ex]{$^{\mbox{$\fnsymbol{CBtnc}$}}$}}}]\addtocounter{%
CBtnc}{1}E-mail address: mckellar@physics.unimelb.edu.au.%
\addtocounter{CBcit}{1}\item[\hfill$^{\arabic{CBcit}}$]%
\renewcommand\theCscr{\arabic{CBcit}}\protect\refstepcounter{Csc%
r}\protect\label{cit:Costella1994}J.~P.~Costella, \renewcommand
\theCscr{Costella}\protect\refstepcounter{Cscr}\protect\label{au%
:Costella1994}\renewcommand\theCscr{1994}\protect\refstepcounter
{Cscr}\protect\label{yr:Costella1994}``Single-Particle Electrody%
namics'', Ph.D.\ thesis, The University of Melbourne (1994), unp%
ublished; available by anonymous-ftp from ftp.ph.unimelb.edu.au,
directory pub/theses/costella; WWW:~http:/$\!$/www.ph.unimelb.ed%
u.au/$\sim$jpc/homepage.htm.\addtocounter{CBcit}{1}\item[\hfill$%
^{\arabic{CBcit}}$]\renewcommand\theCscr{\arabic{CBcit}}\protect
\refstepcounter{Cscr}\protect\label{cit:Newton1949}T.~D.~Newton
and E.~P.~Wigner, \renewcommand\theCscr{Newton and Wigner}%
\protect\refstepcounter{Cscr}\protect\label{au:Newton1949}%
\renewcommand\theCscr{1949}\protect\refstepcounter{Cscr}\protect
\label{yr:Newton1949}``Localized states for elementary systems''%
, Rev.\ Mod.\ Phys. {\bf21}, 400--406 (1949).\addtocounter{CBcit%
}{1}\item[\hfill$^{\arabic{CBcit}}$]\renewcommand\theCscr{\arabic
{CBcit}}\protect\refstepcounter{Cscr}\protect\label{cit:Foldy195%
0}L.~L.~Foldy and S.~A.~Wouthuysen, \renewcommand\theCscr{Foldy
and Wouthuysen}\protect\refstepcounter{Cscr}\protect\label{au:Fo%
ldy1950}\renewcommand\theCscr{1950}\protect\refstepcounter{Cscr}%
\protect\label{yr:Foldy1950}``On the Dirac theory of spin $1/2$
particles and its non-relativistic limit'', Phys.\ Rev. {\bf78},
29--36 (1950).\addtocounter{CBcit}{1}\item[\hfill$^{\arabic{CBci%
t}}$]\renewcommand\theCscr{\arabic{CBcit}}\protect\refstepcounter
{Cscr}\protect\label{cit:Case1954}K.~M.~Case, \renewcommand
\theCscr{Case}\protect\refstepcounter{Cscr}\protect\label{au:Cas%
e1954}\renewcommand\theCscr{1954}\protect\refstepcounter{Cscr}%
\protect\label{yr:Case1954}``Some generalizations of the Foldy--%
Wouthuysen transformation'', Phys.\ Rev. {\bf95}, 1323--1328 (19%
54).\addtocounter{CBcit}{1}\item[\hfill$^{\arabic{CBcit}}$]%
\renewcommand\theCscr{\arabic{CBcit}}\protect\refstepcounter{Csc%
r}\protect\label{cit:Pryce1935}M.~H.~L.~Pryce, \renewcommand
\theCscr{Pryce}\protect\refstepcounter{Cscr}\protect\label{au:Pr%
yce1935}\renewcommand\theCscr{1935}\protect\refstepcounter{Cscr}%
\protect\label{yr:Pryce1935}``Commuting co-ordinates in the new
field theory'', Proc.\ R.\ Soc.\ London~A {\bf150}, 166--172 (19%
35).\addtocounter{CBcit}{1}\item[\hfill$^{\arabic{CBcit}}$]%
\renewcommand\theCscr{\arabic{CBcit}}\protect\refstepcounter{Csc%
r}\protect\label{cit:Pryce1948}M.~H.~L.~Pryce, \renewcommand
\theCscr{Pryce}\protect\refstepcounter{Cscr}\protect\label{au:Pr%
yce1948}\renewcommand\theCscr{1948}\protect\refstepcounter{Cscr}%
\protect\label{yr:Pryce1948}``The mass-centre in the restricted
theory of relativity and its connexion with the quantum theory o%
f elementary particles'', Proc.\ R.\ Soc.\ London~A {\bf195}, 62%
--81 (1948).\addtocounter{CBcit}{1}\item[\hfill$^{\arabic{CBcit}%
}$]\renewcommand\theCscr{\arabic{CBcit}}\protect\refstepcounter{%
Cscr}\protect\label{cit:Dirac1958}P.~A.~M.~Dirac, \renewcommand
\theCscr{Dirac}\protect\refstepcounter{Cscr}\protect\label{au:Di%
rac1958}\renewcommand\theCscr{1958}\protect\refstepcounter{Cscr}%
\protect\label{yr:Dirac1958}\mbox{}\protect\/{\protect\em The Pr%
inciples of Quantum Mechanics\protect\/}, 4th~ed. (Oxford Univer%
sity Press, Oxford, 1958).\addtocounter{CBcit}{1}\item[\hfill$^{%
\arabic{CBcit}}$]\renewcommand\theCscr{\arabic{CBcit}}\protect
\refstepcounter{Cscr}\protect\label{cit:Bjorken1964}J.~D.~Bjorke%
n and S.~D.~Drell, \renewcommand\theCscr{Bjorken and Drell}%
\protect\refstepcounter{Cscr}\protect\label{au:Bjorken1964}%
\renewcommand\theCscr{1964}\protect\refstepcounter{Cscr}\protect
\label{yr:Bjorken1964}\mbox{}\protect\/{\protect\em Relativistic
Quantum Mechanics\protect\/} (Mc\-Graw-Hill, New York, 1964).%
\addtocounter{CBcit}{1}\item[\hfill$^{\arabic{CBcit}}$]%
\renewcommand\theCscr{\arabic{CBcit}}\protect\refstepcounter{Csc%
r}\protect\label{cit:Foldy1951}L.~L.~Foldy, \renewcommand\theCscr
{Foldy}\protect\refstepcounter{Cscr}\protect\label{au:Foldy1951}%
\renewcommand\theCscr{1951}\protect\refstepcounter{Cscr}\protect
\label{yr:Foldy1951}``The electron--neutron interaction'', Phys.%
\ Rev. {\bf83}, 688--688 (1951).\addtocounter{CBcit}{1}\item[%
\hfill$^{\arabic{CBcit}}$]\renewcommand\theCscr{\arabic{CBcit}}%
\protect\refstepcounter{Cscr}\protect\label{cit:Foldy1952a}L.~L.~%
Foldy, \renewcommand\theCscr{Foldy}\protect\refstepcounter{Cscr%
}\protect\label{au:Foldy1952a}\renewcommand\theCscr{1952a}%
\protect\refstepcounter{Cscr}\protect\label{yr:Foldy1952a}``The
electromagnetic properties of Dirac particles'', Phys.\ Rev. {\bf
87}, 688--693 (1952).\addtocounter{CBcit}{1}\item[\hfill$^{%
\arabic{CBcit}}$]\renewcommand\theCscr{\arabic{CBcit}}\protect
\refstepcounter{Cscr}\protect\label{cit:Foldy1952b}L.~L.~Foldy,
\renewcommand\theCscr{Foldy}\protect\refstepcounter{Cscr}\protect
\label{au:Foldy1952b}\renewcommand\theCscr{1952b}\protect
\refstepcounter{Cscr}\protect\label{yr:Foldy1952b}``The electron%
--neutron interaction'', Phys.\ Rev. {\bf87}, 693--696 (1952).%
\addtocounter{CBcit}{1}\item[\hfill$^{\arabic{CBcit}}$]%
\renewcommand\theCscr{\arabic{CBcit}}\protect\refstepcounter{Csc%
r}\protect\label{cit:Foldy1958}L.~L.~Foldy, \renewcommand\theCscr
{Foldy}\protect\refstepcounter{Cscr}\protect\label{au:Foldy1958}%
\renewcommand\theCscr{1958}\protect\refstepcounter{Cscr}\protect
\label{yr:Foldy1958}``Neutron--electron interaction'', Rev.\ Mod%
.\ Phys. {\bf30}, 471--481 (1958).\renewcommand\theCscr{\arabic{%
CBcit}}\protect\refstepcounter{Cscr}\protect\label{citlast}%
\settowidth\Cscr{$^{\ref{cit:Foldy1958}}$}\end{list}\par\end{doc%
ument}
%